\documentclass[]{Format/spie}  

\usepackage{amsmath,amsfonts,amssymb}
\usepackage{graphicx}
\usepackage[colorlinks=true, allcolors=blue]{hyperref}

\title{Spatial frequency response and sensitivity of the nonlinear curvature wavefront sensor}

\author[a]{Stanimir Letchev}
\author[a]{Jonathan Crass}
\author[a]{Justin R. Crepp}
\author[a]{Sam Potier}
\affil[a]{University of Notre Dame, Department of Physics and Astronomy, 225 Nieuwland Science Hall, Notre Dame, Indiana, United States, 46556-5670}

\authorinfo{Further author information: (Send correspondence to S.L.)\\S.L.: E-mail: sletchev@nd.edu}

\begin{document} 
\maketitle

\begin{abstract}
The nonlinear curvature wavefront sensor (nlCWFS) has been shown to be a promising alternative to existing wavefront sensor designs. Theoretical studies indicate that the inherent sensitivity of this device could offer up to a factor of 10$\times$ improvement compared to the widely-used Shack-Hartmann wavefront sensor (SHWFS). The nominal nlCWFS design assumes the use of four detector measurement planes in a symmetric configuration centered around an optical system pupil plane. However, the exact arrangement of these planes can potentially be optimized to improve aberration sensitivity, and minimize the number of iterations involved in the wavefront reconstruction process, and therefore reduce latency. We present a systematic exploration of the parameter space for optimizing the nlCWFS design. Using a suite of simulation tools, we study the effects of measurement plane position on the performance of the nlCWFS and detector pixel sampling. A variety of seeing conditions are explored, assuming Kolmogorov turbulence. Results are presented in terms of residual wavefront error following reconstruction as well as the number of iterations required for solution convergence. Alternative designs to the symmetric four-plane design are studied, including three-plane and five-plane configurations. Finally, we perform a preliminary investigation of the effects of broadband illumination on sensor performance relevant to astronomy and other applications.  

\end{abstract}


\section{INTRODUCTION}
\label{sec:intro}

The design of any wavefront sensor (WFS) involves a fundamental trade-off between sensitivity, dynamic range, and speed. Challenging applications for adaptive optics (AO) systems, such as high-contrast imaging, precision radial velocity measurements using single-mode fibers, laser communications, remote sensing, and space domain awareness, have requirements that demand high sensitivity (ability to correct at low photon levels), a large dynamic range for aberration correction (phase errors of many waves), and fast operating speeds (many kHz) simultaneously. A nonlinear version of the curvature wavefront sensor (hereafter referred to as the nlCWFS) has been proposed to address these challenges.\cite{guyon_10,mateen_11,Crepp:20} The nlCWFS differs from the conventional curvature WFS by incorporating four (rather than two) intensity measurement planes that are defocused from the pupil in order to sense both high and low spatial frequencies. The need for more than two planes is a consequence of the Talbot effect, which quantifies the characteristic length scale over which monochromatic interference fringes repeat as a coherent beam propagates. The Talbot distance, $Z_T$, is defined,
\begin{equation}
  Z_T=\frac{2 a^2}{\lambda},
  \label{eq_talbot_dist}
\end{equation}
where $a$ represents the spatial period of a periodic disturbance and $\lambda$ is wavelength \cite{wen_13}. Since $Z_T$ depends on $a$, the optimal defocus distances depend on the power spectrum of aberrations experienced by the optical system.

Previous analyses of the nlCWFS have, by default, assumed a sensor design with four detector planes, with two pairs located symmetrically on either side of the pupil plane, but this configuration is not strictly required. In addition, previous studies have typically determined optimal measurement plane distances through trial and error based on experimental results, dependent on each individual system. There has not yet been a more general solution to answer the question of what are the ideal distances (if any) for placement of the nlCWFS detector planes.

The nlCWFS has also previously been treated in a mostly monochromatic context. However, astronomy-focused applications of AO frequently use natural guide stars as their reference object. The blackbody emission of a distant source requires a WFS capable of using broadband light in order to have sufficient flux for wavefront reconstruction. In addition, a wide bandpass is necessary when observing faint sources for adequate WFS speed and performance.

These proceedings describe our analysis of the nlCWFS in the context of optimizing its use for a broad range of applications.  In \S \ref{sec:methods}, we describe the general numerical methods used for simulations. In \S \ref{sec:kolmogorov}, we explore the optimal distances of the four-plane configuration and study alternative sensor configurations (three and five planes). In \S \ref{sec:sampling}, we study the effects of detector spatial sampling while in \S \ref{sec:broadband}, we model the effects of broadband illumination. Finally, conclusions are provided in \S \ref{sec:conclusions}.
\section{Numerical Methods}\label{sec:methods}


Due to the nonlinear optical propagation regime in which it operates, the nlCWFS must use an iterative reconstruction method to recover wavefront phase. The approach to extracting phase and amplitude information for this sensor is often based on the Gerchberg-Saxton (GS) method, where light is numerically propagated between multiple measurement planes and the optical system pupil \cite{guyon_10}. This method of recording intensity information from (typically) four measurement planes allows for an attempt to reconstruct the complex field at the pupil. Figure \ref{fig:plane_config} shows the location of measurement planes displaced from the pupil along a collimated beam, which is typically generated using a reimaging relay.


Because of the nonlinearity of the nlCWFS and the large parameter space of plausible configurations (defocus distances, number of planes, spatial sampling, etc.), it is difficult to determine its performance characteristics analytically. Instead, we have developed a series of simulations to model the sensor's performance numerically. Custom MATLAB scripts were developed to model, sense, and reconstruct near-field (Fresnel) diffraction effects. These programs were further supplemented by commercially- and publicly-available MATLAB codes (see below). The combined tools were used to generate simulated data and also as an integral part of the wavefront reconstruction algorithm. 



\subsection{Propagation Method}\label{sec:prop_method}

Since the nlCWFS's image generation process (\S \ref{sec:img_gen}) and reconstruction algorithm (\S \ref{sec:GS_alg}) rely primarily on electric field propagation, it was necessary to choose a robust propagation algorithm for both simulations and phase recovery. Due to the distances, wavelength, and beam diameters used, all simulations were performed within the Fresnel (near-field) regime for coherent light propagation. The propagation algorithms selected were based on the angular spectrum code of Schmidt 2010\cite{schmidt_10}. This code was slightly modified for efficiency, as some variables, such as the quadratic phase factors, did not need recalculation with each iterative loop.

\subsection{Phase Aberrations and Image Generation}\label{sec:img_gen}

\begin{figure*}
    \centering
    \includegraphics[width=\textwidth]{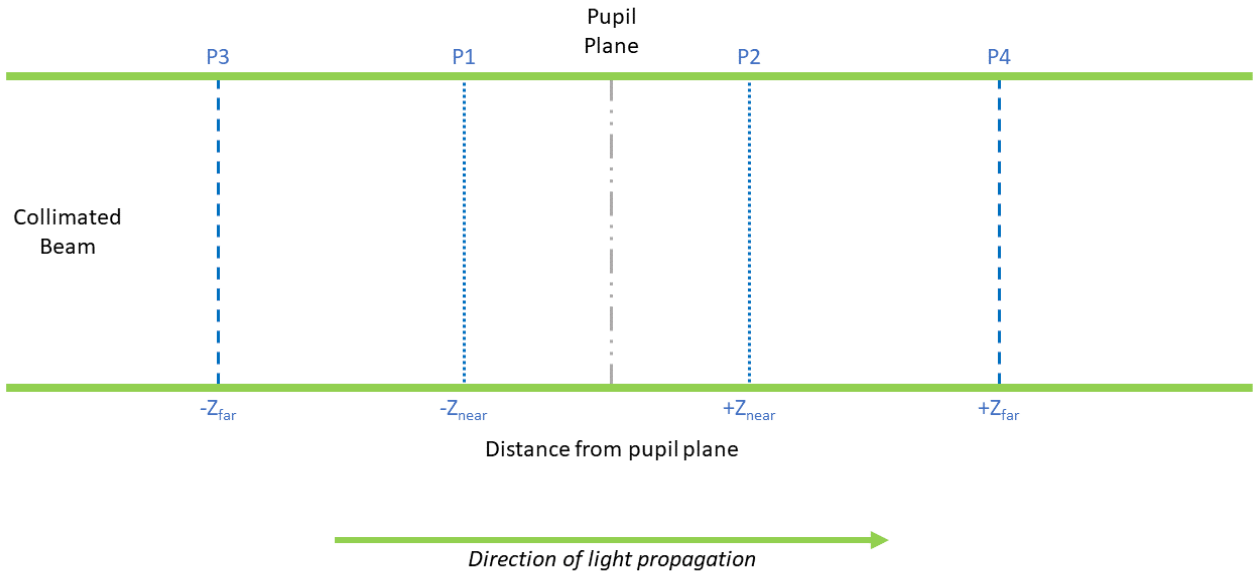}
    \caption{Visual representation of plane configuration including names and definitions.}
    \label{fig:plane_config}
\end{figure*}

Phase aberrations used to simulate atmospheric turbulence disturbances at the system pupil plane were created using the WaveProp \textit{kolmogphzscreen} function from the WaveProp simulation package.\cite{Brennan:16,Brennan:17} The phase masks corresponding to these aberrations were kept for comparison to ensure that residual phase errors were calculated based on the original phase values rather than on extracting a phase from the total electric field, as this would be subject to phase unwrapping errors. Pupil-plane phase masks were then padded with zeros to ensure sufficient sampling in Fourier space.

The pupil-plane electric field was created by combining the generated phase masks with a uniform electric field amplitude. This electric field was then propagated to each measurement plane using the method outlined in $\S$\ref{sec:prop_method}. The planes were defined using the same notation as in Crass et al. 2014\cite{crass_14}. For the symmetric four-plane configuration, the main parameters that define the sensor are the $Z_{\rm near}$ and $Z_{\rm far}$ distances, that signify how far away from the pupil the near and far planes are, respectively (Figure \ref{fig:plane_config}). The simulated intensities at each plane, analogous to detector measurements in a real-world nlCWFS, were then used as input for the wavefront reconstruction process.

\subsection{Modified Gerchberg-Saxton Algorithm}\label{sec:GS_alg}

\begin{figure*}
    \centering
    \includegraphics[width=\textwidth]{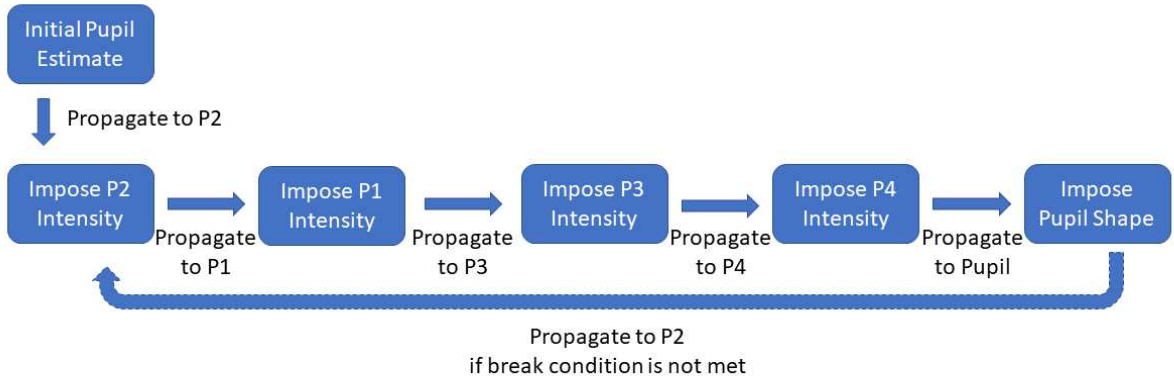}
    \caption{Visual representation of the modified GS algorithm used for wavefront reconstruction, showing the propagation order and steps. The dashed arrow represents where the loop would break when the convergence criterion is met. Figure adapted from Guyon 2009.\cite{guyon_10}}
    \label{fig:GS_Algorithm}
\end{figure*}

The reconstruction method used for simulating the nlCWFS is similar to the modified GS algorithm outlined in Guyon 2009 and shown in Figure \ref{fig:GS_Algorithm}\cite{guyon_10}. The electric field is propagated from one measurement plane to another. The intensity ``measurement'' at each location is imposed onto the field by replacing the electric field amplitude with the square-root of the intensity. The electric field is eventually propagated back to the pupil. Field points located outside of the pupil are set to zero to minimize numerical noise, since no light should arise from outside of that boundary. 

As the algorithm is iterative, a criterion for when it has reached a solution is required. In our simulations, the loop's convergence condition was defined as a lower limit to the RMS of the difference, or residual, in wavefront phase between consecutive reconstruction loops. For situations in which the wavefront did not meet the convergence criterion, a maximum number of reconstruction loops was specified to reduce simulation time.

\subsection{Phase Unwrapping}\label{sec:phase_unwrapping}

A consequence of the GS reconstruction method is that the phase can only be extracted in a range of $\pm \lambda/2$, causing discontinuities for large wavefront aberrations. This effect can inhibit wavefront correction when using continuous deformable mirror surfaces. To provide consistent phase unwrapping, we chose an advanced phase unwrapping method called ``\textit{xphase}'' from the WaveProp package\cite{Brennan:16,Brennan:17}. We find that \textit{xphase} consistently provides more accurate phase unwrapping results compared to other methods tested, and therefore, all simulations use this method.


\section{Defocus Distance and Plane Configuration Optimization}\label{sec:kolmogorov}

To study the ideal number and location of detector planes, we performed simulations of the nlCWFS using phase screens resembling atmospheric turbulence. These tests varied plane location for different atmospheric turbulence strengths (\S \ref{sec:kolmogorov_dr0_results}), as well as number of planes (\S \ref{sec:kolmogorov_threeplane_results} and \ref{sec:kolmogorov_fiveplane_results}). The general simulation outline and parameters are described in Section \ref{sec:komolgorov_simparam}.

\subsection{Simulation Parameters}\label{sec:komolgorov_simparam}
To identify ideal measurement plane distances, a series of 16 randomly-seeded Kolmogorov input wavefront aberrations were tested over a wide range of detector plane distances. A separate series of 16 wavefronts were created for each global turbulence parameter. Detector plane distances were chosen to pass through the Talbot distance corresponding to the spatial size of the Fried parameter ($r_0$) of the turbulence, and were moved from the pupil to greater distances to investigate if any diffraction effects manifested at longer distances. Simulations used a pupil diameter of $D=0.5$ m and monochromatic light at wavelength $\lambda=532$ nm. 

Spatial sampling was set to correspond to a 64-pixel-diameter pupil for all plane configurations, which ensured sufficient sampling to not limit reconstruction accuracy (see \S \ref{sec:sampling} for further analysis). For all reconstructions, the maximum number of reconstruction loops was set to 20. The wavefront reconstruction convergence criterion was set conservatively to 1 nm, as preliminary tests showed this produced valid results. Relevant performance metrics (RMS of the residual wavefront error (WFE), number of iterations needed for convergence, time to complete reconstruction, etc.) were recorded for each combination. These values were then averaged over all 16 simulated wavefronts for each variation of input parameters, to provide more statistical validity to the results.

\subsection{Optimal Detector Plane Analysis}\label{sec:kolmogorov_dr0_results}

\begin{figure}
    \centering
    \includegraphics[width=\textwidth]{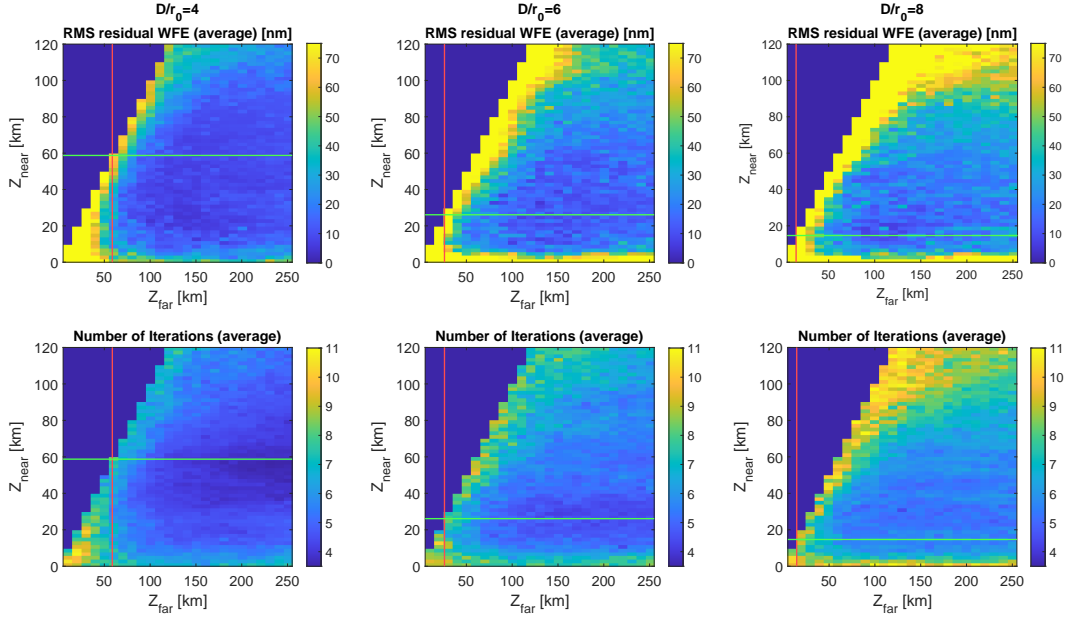}
    \caption{Symmetric four-plane reconstruction of Kolmogorov phase aberrations, showing the average values of the RMS of the residual WFE (top row) and the number of iterations (bottom row). The distance corresponding to the Talbot distance of each $r_0$ value are shown as horizontal green lines for $Z_{\rm near}$ and as vertical red lines for $Z_{\rm far}$. Note that the top-left of the plot has no data, as this region is where $Z_{\rm near}$ is larger than $Z_{\rm far}$. Since this is symmetrically the same as those values being switched, this region was set to NaN values to reduce simulation time.}
    \label{fig:4_plane_kolmogorov_Dr0}
\end{figure}

Due to the Talbot effect, the defocus distance at which phase aberrations manifest as intensity aberrations is closely linked to the spatial frequency of an aberration in the pupil plane. For Kolmogorov turbulence, the Fried parameter, $r_0$, characterizes the scale of aberrations. As such, it was a relevant metric for the investigation into ideal defocus parameters. We use the ratio $D/r_0$ to characterize the effects of turbulence on defocus distance, as it is a unitless metric that can be scaled for other systems. To see the effect of $D/r_0$ on the ideal detector placement of the nlCWFS, the symmetric four-plane configuration was tested using three different values of $D/r_0$ (4, 6, and 8) and run through a broad parameter space of distances.

The results for this analysis are shown in Figure \ref{fig:4_plane_kolmogorov_Dr0}. For all $D/r_0$ values, it can be seen that, at very short $Z_{\rm near}$ distances, and at distances where $Z_{\rm near}$ and $Z_{\rm far}$ are very close together (the bottom and top-left of the plots, respectively), there is a large increase in the RMS residual WFE (over 200 nm) and the number of iterations (over 10), signifying a poor reconstruction. The large increase when the planes are too close to the pupil is because any substantial diffraction has yet to occur, giving the sensor very little information about the pupil phase. The poor performance when the planes are too close together is due to an effective repeat of information, i.e. the planes are so similar to each other that the sensor is essentially acting as a two-plane configuration.

The best reconstructions occur when both the RMS residual WFE and the number of iterations are low. This means that the reconstructor produced a very accurate reconstruction in a short amount of time, which is the ideal case for many AO applications. The lowest values tended to occur in a pattern that looks like a ``valley'' of small values, primarily in the number of iterations, but also faintly in RMS residual WFE. The values of RMS residual WFE in this low region were around 5 to 10 nm and the number of iterations was around 3 to 5. This valley can be seen around the Talbot distance corresponding to an aberration with a period equal to the spatial size of $r_0$, for the $Z_{\rm near}$ plane distances (shown as a green line in Figure \ref{fig:4_plane_kolmogorov_Dr0}). As $D/r_0$ increases, the valley feature occurs closer to the pupil in $Z_{\rm near}$ (and partially in $Z_{\rm far}$), as expected due to the Talbot effect. 


The minimum optimal $Z_{\rm far}$ value did not easily map to any multiple of Talbot distance, for $r_0$ or the telescope diameter. For the values tested in these simulations, the minimum optimal $Z_{\rm far}$ was seen to occur at around 100 to 200 km. To scale these values appropriately to a differently-sized telescope, the Fresnel number ($F=D^2 Z/\lambda$) can be used. If the Fresnel number is conserved, the correct minimum $Z_{\rm far}$ distance can be found for any size pupil diameter and any observing wavelength. This hypothesis was tested with our simulations and found to be accurate.

The results show that the optimal defocus distances encompass a wide range of plausible values. Primarily, as long as two factors are met, the sensor appears to show optimal or near-optimal performance (low RMS residual WFE and low number of iterations):

\begin{enumerate}
    \item The $Z_{\rm near}$ value is close to the Talbot distance corresponding to the spatial size of $r_0$, and
    \item The $Z_{\rm far}$ value is \textit{at least} a value larger than approximately 100 - 200 km for a pupil with $D=0.5$ m and wavelength $\lambda=532$ nm. This value can be scaled to other pupil diameters and wavelengths by conserving Fresnel number.
\end{enumerate}

\subsection{Three-Plane Design}\label{sec:kolmogorov_threeplane_results}

\begin{figure}
    \centering
    \includegraphics[width=\textwidth]{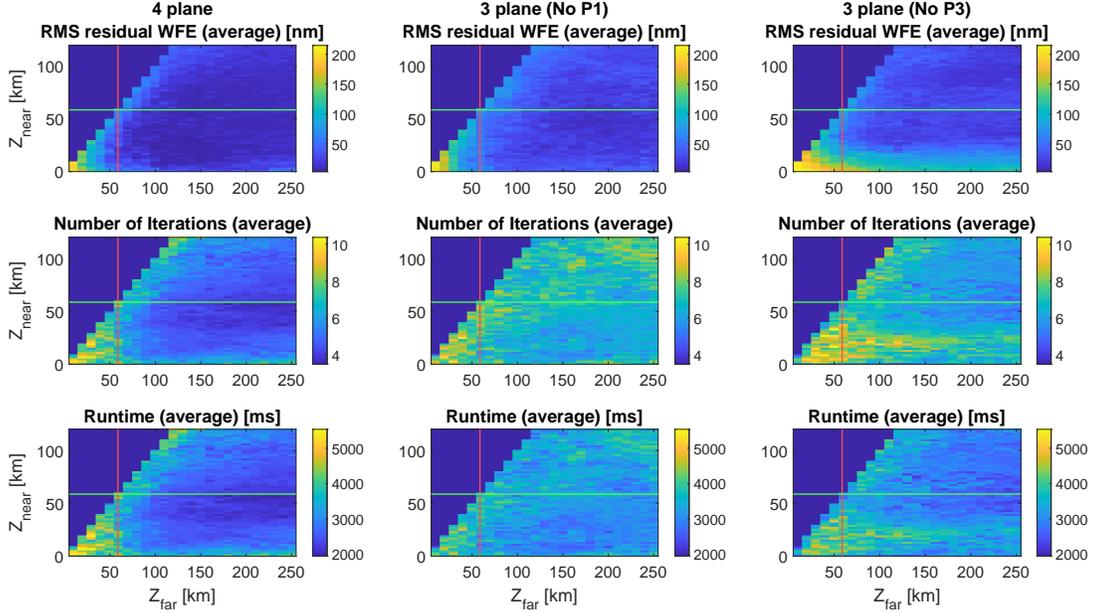}
    \caption{Three-plane and four-plane reconstructions of a Kolmogorov phase aberration with a $D/r_0=4$, showing the average values of the RMS of the residual WFE (top row), the average number of iterations (middle row), and the approximate average reconstruction time based on the average time per loop (bottom row). The distance corresponding to the Talbot distance of each $r_0$ value are shown as horizontal green lines for $Z_{\rm near}$ and as vertical red lines for $Z_{\rm far}$.}
    \label{fig:3_plane_kolmogorov}
\end{figure}

It is relevant to look at alternative numbers of plane designs, as fewer planes reduce the number of Fourier transforms needed per iteration in the reconstruction and thus may allow for faster reconstruction times. In addition, fewer planes allow for a more compact optical design and increased flux per plane as fewer splitting optics are required. To test the performance of a three-plane configuration, the same plane definitions were used as those for the symmetric four-plane design. The configuration was identical except one plane was removed, and only $D/r_0=4$ was evaluated. For one set of simulations, the P3 ($-Z_{\rm far}$) plane was removed, and for the second simulation, the P1 ($-Z_{\rm near}$) plane was removed.

Results for these two sets of simulations are shown in Figure \ref{fig:3_plane_kolmogorov}. It can be seen that both three-plane configurations produce RMS residual WFE values comparable to the symmetric four-plane configuration. However, both three-plane configurations require more iterations to reach the same RMS residual WFE values. Despite a shorter time per iteration (a three-plane loop took approximately 88\% of the time of a four-plane loop), the increased number of iterations compared to the four-plane configuration requires an increased total time overall to recover the wavefront.

\subsection{Five-Plane Design}\label{sec:kolmogorov_fiveplane_results}

\begin{figure*}
    \centering
    \includegraphics[width=\textwidth]{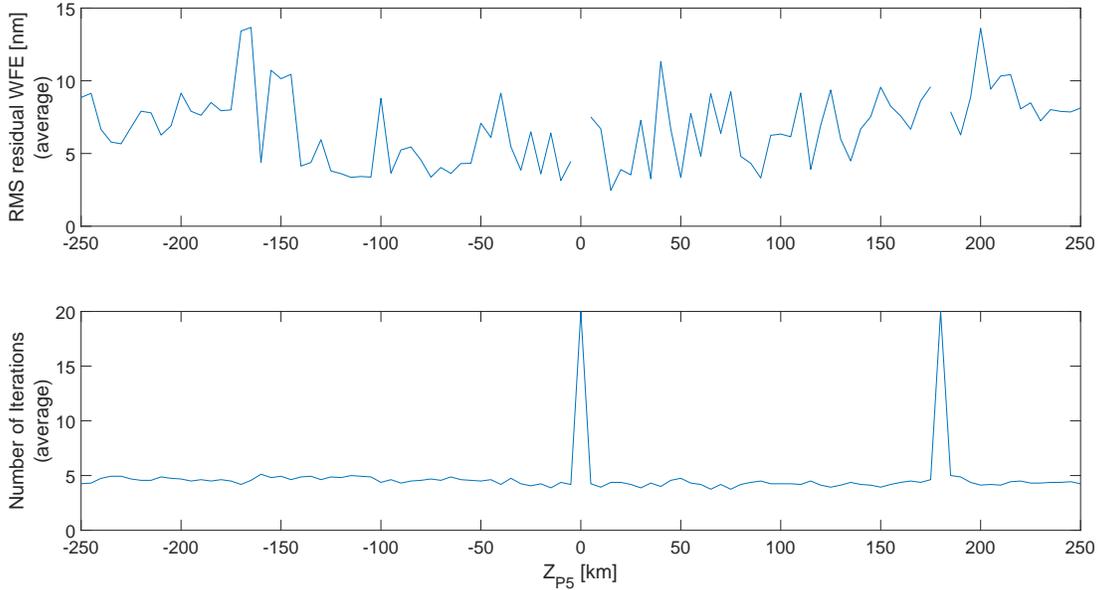}
    \caption{Five-plane reconstructions of a Kolmogorov phase aberration with $D/r_0 = 4$, showing average values of the RMS of wavefront residuals (top) and the average number of iterations required to reach a 1 nm convergence criterion (bottom). Gaps in the RMS plot and peaks in the iteration plot correspond to zero-distance propagations (i.e. when P5 was placed at the location of either P4 or the pupil). Since the P5 plane was placed in between P4 and the pupil in the modified GS loop, these locations would result in the algorithm attempting to propagate at a zero distance, and results in numerical errors that inflate the residual RMS and number of iterations.}
    \label{fig:5_plane_kolmogorov}
\end{figure*}

The results of the three-plane design suggested that adding additional planes may provide a reconstruction time advantage and therefore a design with five planes was studied. Adding a fifth plane, while continuing to vary the $Z_{\rm near}$ and $Z_{\rm far}$ distances, would be time consuming to simulate and challenging to analyze, so we instead chose the best $Z_{\rm near}$ and $Z_{\rm far}$ values from the four-plane results and ran a single extra plane through the same range of values as before (Z = -250 to 250 km). The set of parameters used was $Z_{\rm near}=47$ km and $Z_{\rm far}=180$ km with a $D/r_0=4$. The results are shown in Figure \ref{fig:5_plane_kolmogorov}. 

We find that the average RMS residual WFE varies between $\approx$ 3 nm and 14 nm, and average number of iterations typically range between 4 and 5. The symmetric-four-plane values at those same distances were 6.5 nm and 4.2 iterations, respectively. This result indicates that the five-plane configuration does not offer much, if any, improvement in RMS residual WFE compared to the four-plane design, while also not offering any improvement in iteration number, causing the overall reconstruction time to increase. This combination of factors indicate that the four-plane design appears to be the best overall choice.
\section{Spatial Sampling Requirements}\label{sec:sampling}

The number of pixels used to sense light across the beam impacts WFS performance in two ways:
\begin{enumerate}
    \item Wavefront recovery requires a threshold (minimum) number of pixel sampling for accurate results;
    \item Increased pixel usage increases processing needs, resulting in longer computation time and, therefore, increased system latency.
\end{enumerate}
For the nlCWFS, the requirements on spatial sampling are somewhat more complicated than conventional sensors because, by definition, the planes of observation are not located at a pupil or focal plane. 

We study how spatial sampling impacts accuracy of the reconstruction process by systematically adjusting the number of samples relative to the value of $r_0$ across the system pupil. This was done using a similar method to the defocus distance testing and followed the example of Banet et al. 2017 for digital holography.\cite{banet_17} A separate series of 16 wavefronts were created for $D/r_0$ values of 4, 6, 8, and 10, for which the results were averaged for statistical relevance. These wavefronts were then reconstructed using a symmetric-four-plane nlCWFS configuration with measurement planes located at the optimal distances derived from the analysis described in Section \ref{sec:kolmogorov_dr0_results}.

Starting with a high sampling rate of 128 pixels across the beam diameter, both the initial wavefront and measurement plane intensities were then binned by increasing factors (1, 2, 4, 8, and 16). We define our spatial sampling in terms of the size of $r_0$ in pixels, as evaluated in the pupil plane. Since the simulations use identical spatial scaling for all planes, this definition also establishes the spatial sampling in the measurement planes. The field-estimated Strehl ratio ($S_F$) was used as the main metric for performance, so as to easily compare results to the Banet et al. 2017 digital holography analysis \cite{banet_17}. This metric also helps to avoid phase unwrapping effects which could impact the RMS residual WFE more strongly than the field-estimated Strehl ratio.



\begin{figure*}
    \centering
    \includegraphics[width=0.75\textwidth]{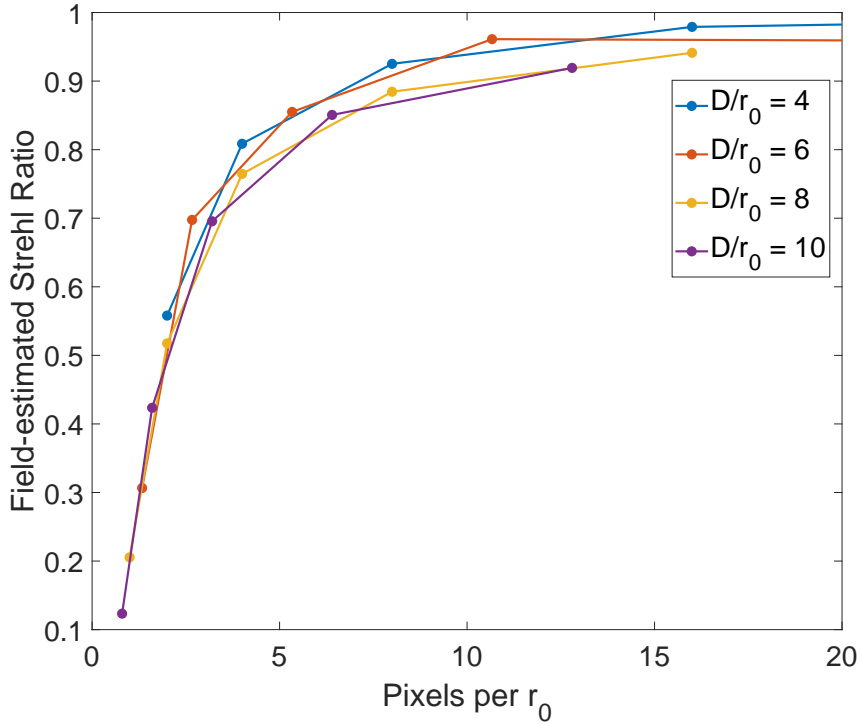}
    \caption{The impact of spatial sampling on field-estimated Strehl ratio as a function of turbulence strength.}
    \label{fig:sampling_plot}
\end{figure*}

The results of this analysis are shown in Figure \ref{fig:sampling_plot}. We find that the measurement planes must be sampled at greater than or equal to approximately 4 to 5 pixels per $r_0$ for a reconstruction that produces a field-estimated Strehl ratio sufficient for diffraction-limited performance ($S_F\approx 0.8$). All other simulations in our analysis intentionally exceeded this value.
\section{Broadband Analysis}\label{sec:broadband}

To study the effects of broadband illumination, effective broadband images were generated at the individual measurement planes and used as the input for wavefront reconstruction. To generate images in broadband light, we defined a bandpass for each of the four defocus planes and then propagated a series of monochromatic images at individual wavelengths through the bandpass to the measurement plane. These were then averaged together to create the broadband image.

Broadband analysis used a spatial sampling corresponding to 128 pixels across a $D=0.5$ m pupil, $D/r_0=6$, and defocus distances (at 532 nm) of $Z_{\rm near}=47$ km and $Z_{\rm far}=180$ km. The central wavelength of each bandpass was used to modify the defocus distance of broadband images through conservation of Fresnel number from the 532 nm distances allowing the reconstructor to work as if the input light was monochromatic at 532 nm. Bandpass ranges and wavelength sampling are shown in Table \ref{tab:bandpass_values}.

\begin{table}[ht]
\caption{Bandpass parameters for the nlCWFS broadband analysis} 
\label{tab:bandpass_values}
\begin{center}       
\begin{tabular}{|l|c|c|c|} 
\hline
\rule[-1ex]{0pt}{3.5ex}  Band & narrow & medium & wide \\
\hline
\rule[-1ex]{0pt}{3.5ex}  P1 band [nm] & 500 - 549 & 400 - 499 & 300 - 499  \\
\hline
\rule[-1ex]{0pt}{3.5ex}  P1 central wvl [nm] & 525 & 450 & 400   \\
\hline
\rule[-1ex]{0pt}{3.5ex}  P2 band [nm] & 550 - 599 & 500 - 599 & 500 - 69   \\
\hline
\rule[-1ex]{0pt}{3.5ex}  P2 central wvl [nm] & 575 & 550 & 600   \\
\hline
\rule[-1ex]{0pt}{3.5ex}  P3 band [nm] & 600 - 649 & 600 - 699 & 700 - 899  \\
\hline
\rule[-1ex]{0pt}{3.5ex}  P3 central wvl [nm] & 625 & 650 & 800  \\
\hline
\rule[-1ex]{0pt}{3.5ex}  P4 band [nm] & 650 - 699 & 700 - 799 & 900 - 1099 \\
\hline
\rule[-1ex]{0pt}{3.5ex}  P4 central wvl [nm] & 675 & 750 & 1000  \\
\hline
\rule[-1ex]{0pt}{3.5ex}  Wvl Sampling [num of images per plane] & 25 & 20 & 40 \\
\hline
\end{tabular}
\end{center}
\end{table} 

\begin{figure}
    \centering
    \includegraphics[width=\textwidth]{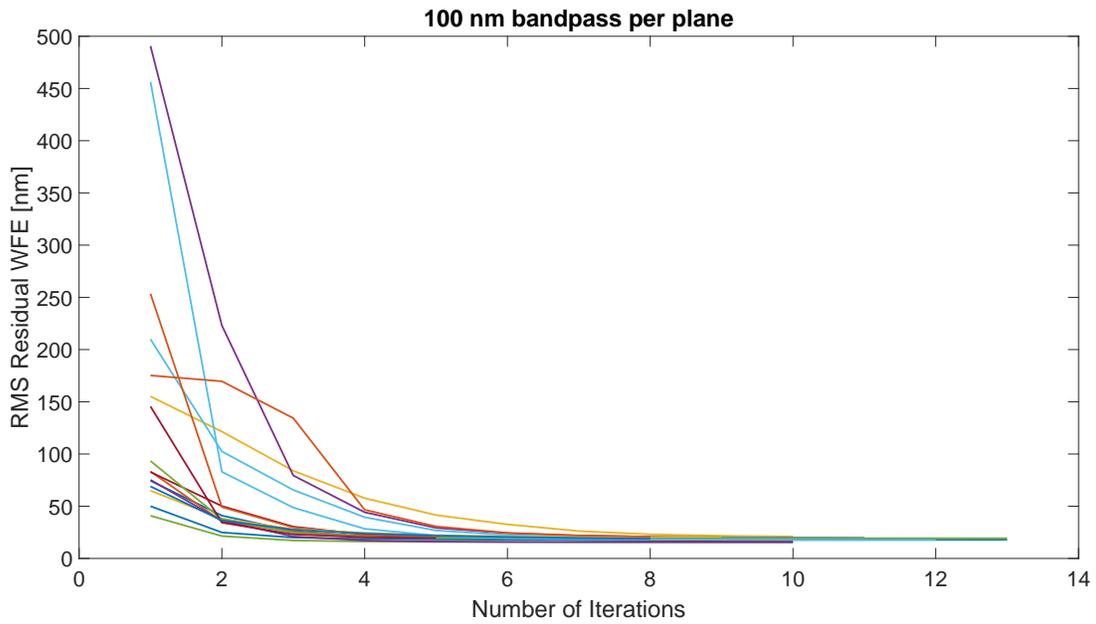}
    \caption{Results of 16 different aberrations using $D/r_0=6$, showing the residual RMS after each loop of the nlCWFS reconstructor. A bandpass of 100 nm was used in each of the four channels. All aberrations reach a lower limit of correction that does not improve with increasing numbers of iterations.}
    \label{fig:broadband_16_runs}
\end{figure}

The main factor analyzed was how broadband light affects the accuracy and speed of the reconstruction process. To test this, the RMS residual WFE was calculated at each iteration step, rather than at the end of the reconstruction loop. The plot for the medium (100 nm) bandpass is shown in Figure \ref{fig:broadband_16_runs}. The RMS residual WFE for all reconstructions has a different starting value depending on the strength of the input aberration, but converges to a minimum quantity by 6-8 iterations.

\begin{figure}
    \centering
    \includegraphics[width=\textwidth]{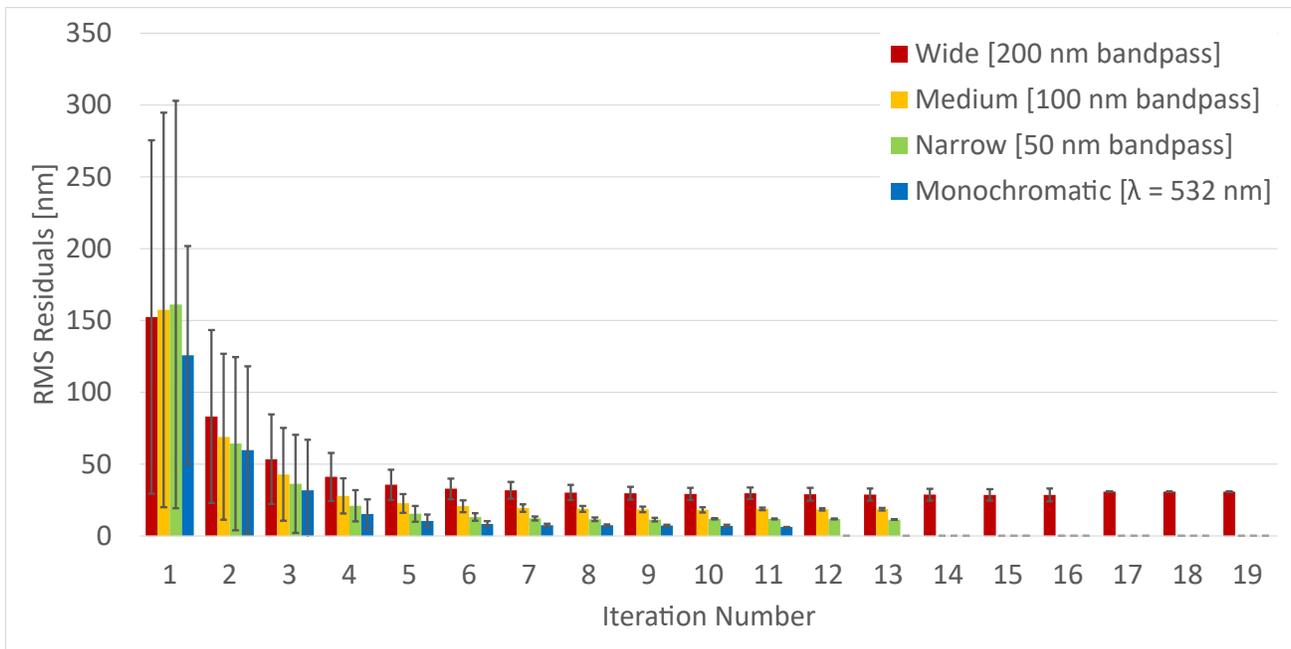}
    \caption{The average and standard deviation of the RMS residual WFE for four sets of data each using a different bandpass. The results show the effect of changing bandpass on the convergence speed of a nlCWFS system. Each band has a lower limit of correction, with wider bandpasses having a higher limit on RMS residual WFE compared to narrower bands. A monochromatic case is shown for reference.}
    \label{fig:broadband_full}
\end{figure}

To better understand the differences between each bandpass (and not plot 64 lines on the same plot), the average and standard deviation of results from Figure \ref{fig:broadband_16_runs} were taken. These are shown in Figure \ref{fig:broadband_full}. As can be seen, the ``floor'' of RMS residuals changes for each bandpass, increasing as the bandpass gets wider. It can also be seen that, in general, the convergence criterion is reached later for wider bandpasses. However, despite this change, the increase of floor does not significantly impact the RMS residuals, remaining below 50 nm even when using a 200 nm bandpass per measurement plane (800 nm in total).

\section{Conclusions}\label{sec:conclusions}

Using a series of numerical simulations, we have studied the optimal configuration of the nlCWFS for astronomy-focused AO applications, with relevance to other fields. This sensor offers both improved sensitivity and dynamic range compared to the industry-standard Shack-Hartmann, but has a tradeoff in speed. With improvements in computing, the impact of this tradeoff can be minimized. Studying the optimal distances of the detector planes, the optimal number of detector planes, and the effects of detector sampling and broadband input allow us to characterize the nlCWFS and improve its performance for practical use. The outcomes of our theoretical study are:

\begin{itemize}
    \item The nlCWFS works best when $Z_{\rm near}$ is close to the Talbot distance corresponding to the spatial size of $r_0$.
    \item The optimal $Z_{\rm far}$ distance is further than a Fresnel-number-scaled equivalent of 100 - 200 km for a $D=0.5$ m aperture with $\lambda =532$ nm.
    \item A three-plane configuration produces acceptable residual RMS but increases overall reconstruction time compared to the symmetric four-plane design.
    \item A five-plane configuration does not seem to offer any significant improvement in residual RMS and increases overall reconstruction time, meaning it is less practical than the four-plane design.
    \item Using at least 4 to 5 pixels per $r_0$ as defined in the pupil plane for each of the four detectors provides sufficient sampling to achieve diffraction-limited performance.
    \item Increasing the bandpass entering each of the four detector planes degrades reconstruction performance in terms of both speed and accuracy, but does not significantly degrade sensor performance.
\end{itemize}

\acknowledgments 

We thank Dr. Mark Spencer for helpful technical discussions. This research was supported in part by the Air Force Research Laboratory AFRL/RD-AFRL-Directed Energy, Kirtland Air Force BaseDirectorate, through the Air Force Office of Scientific Research Summer Faculty Fellowship Program\textregistered, Contract Numbers FA8750-15-3-6003, FA9550-15-0001 and FA9550-20-F-0005. This work was supported in part by the Northrop Grumman Corporation. 

\bibliography{bib/report} 
\bibliographystyle{Format/spiebib} 

\end{document}